\documentclass[11pt]{article}
\usepackage[english]{babel}
\usepackage{amsmath}
\usepackage{amsthm}
\usepackage{amsfonts}
\usepackage{amssymb}
\usepackage{enumerate}
\usepackage{mathtools}
\usepackage{enumitem}
\usepackage{bbm} 
\usepackage{bm}

\DeclarePairedDelimiter\floor{\lfloor}{\rfloor}
\usepackage{fullpage}
\usepackage{authblk}

\DeclareMathOperator*{\argmax}{arg\,max}

\newcommand{\be}{\begin}
\newcommand{\e}{\end}
\newcommand{\beq}{\begin{equation}}
\newcommand{\eeq}{\end{equation}}
\newcommand{\beqs}{\begin{equation*}}
\newcommand{\eeqs}{\end{equation*}}

\renewcommand{\l}{\left}
\renewcommand{\r}{\right}

\newcommand{\set}[1]{\mathbb{#1}}

\newcommand{\curly}[1]{\mathcal{#1}}

\newcommand{\setof}[2]{\left\{ #1\; : \;#2 \right\}}

\newcommand{\C}{\set{C}}

\newcommand{\gam}{\gamma}

\newcommand{\ket}[1]{|#1\rangle}
\newcommand{\bra}[1]{\langle#1|}

\newcommand{\Tr}{\mathrm{Tr}}	

\theoremstyle{definition}

\theoremstyle{remark}

\begin{document}

\title{Entropic relations for\\ indistinguishable quantum particles}
\date{February 20, 2024}

\author{Marius Lemm}

\affil{Department of Mathematics, University of T\"ubingen, 72076 T\"ubingen, Germany}

\maketitle

\begin{abstract}
The von Neumann entropy of a $k$-body reduced density matrix $\gamma_k$ quantifies the entanglement between $k$ quantum particles and the remaining ones. In this short paper, we rigorously prove general properties of this entanglement entropy as a function of $k$: it is concave for all $1\leq k\leq N$ and non-decreasing until the midpoint $k\leq \floor{N/2}$. The results hold for indistinguishable quantum particles and are independent of the statistics.
\end{abstract}

\section{Introduction}
Understanding quantum entanglement lies at the core of quantum theory. It is at the same time a problem of fundamental conceptual importance and of immense practical relevance, in part because quantum entanglement serves as a resource for quantum information-processing protocols such as quantum cryptography, quantum teleportation, superdense coding, and measurement-based quantum computation \cite{bennett1992communication,bennett1996purification,bennett1996mixed,horodecki2009quantum,buscemi2011entanglement,datta2014second,datta2016second}.

A central task of quantum information theory is to investigate and compare various measures of quantum entanglement. A common entanglement measure for pure quantum states is the \textit{entanglement entropy}. Given a pure quantum state $\rho_{AB}$ on a composite Hilbert space $\mathcal H_A\otimes \mathcal H_B$, the entanglement entropy between the subsystems $A$ and $B$ is defined as the von Neumann entropy of the reduced density matrix $\rho_A=\mathrm{Tr}_B[\rho_{AB}]$ \cite{NC}, i.e.,
\[
S(\rho_A)=-\mathrm{Tr}[\rho_A\log\rho_A].
\]
(A pure product state $\rho_{AB}=\rho_A\otimes \rho_B$ then has zero entanglement entropy as it should.) By the Schmidt decomposition one has $S(\rho_A)=S(\rho_B)$, so the entanglement entropy is identical from the perspective of either subsystem.

The entanglement entropy has been extensively studied and used in quantum information theory, high-energy theory, and quantum many-body physics; see, e.g.,  \cite{horodecki2009quantum,calabrese2004entanglement,kitaev2006topological,hayden2006aspects,verstraete2006criticality,banuls2007entanglement,hastings2007area,amico2008entanglement,brandao2008entanglement,nishioka2009holographic,eisert2010colloquium,christandl2012entanglement,brandao2015exponential,landau2015polynomial,gong2017entanglement,cirac2021matrix,anshu2022area,jauslin2022random} and references therein. It has also been measured in experiments \cite{islam2015measuring}. In practice, many different bipartite decompositions of the total Hilbert space into $\mathcal H_A\otimes \mathcal H_B$ are considered. In a quantum many-body system, it is common (see the list of references above) to either take $A$ and $B$ as spatial regions  or to take them as different groups of particles. In the present paper, we focus on the latter perspective.

We consider a system of $N$ indistinguishable quantum particles. (Our results hold for arbitrary particle statistics.) We write $\mathcal H$ for the one-body Hilbert space which we assume is separable. For any number $1\leq k\leq N$, the $N$-particle Hilbert space can be decomposed as $ \mathcal H^{\otimes N}=\mathcal H^{\otimes k}\otimes \mathcal H^{\otimes N-k}$. Given a pure $N$-particle state $\rho_N$, our \textit{main quantity of interest} is the entanglement entropy between the first $k$ particles and the remaining $N-k$ ones, which is defined as
\[
S_k=S(\gam_k)=-\Tr[\gam_k\log \gam_k]
\]
where the \textit{reduced $k$-body density matrix} (or $k$-body quantum marginal) is given by
\[
\gam_k=\Tr_{k+1,\ldots,N}[\rho_N]. 
\]
(We use the convention that the partial trace is trace-preserving and so each $\gamma_k$ is a normalized quantum state with $\Tr[\gam_k]=1$.) The Schmidt decomposition and the indistinguishability of the particles imply the symmetry property
\beq\label{eq:symm}
S_k=S_{N-k},\qquad 1\leq k\leq N-1.
\eeq

The family of $\{S_k\}_{1\leq k\leq N}$ captures the entanglement properties of the underlying physical system, which in particular strongly depends on the particle statistics. Since they depend on delicate many-body properties of the wave function, the $S_k$'s are difficult to compute for most pure states $\rho_N$. As a touchstone, let us briefly consider the $S_k$'s of the following two paradigmatic types of $N$-particle quantum states.
\begin{itemize}
\item For a bosonic product state $\Psi_N=\phi\otimes\phi\otimes\ldots \otimes \phi$, we have $S(\gam_k)=0$ for all $k$. Indeed, the particles are completely independent and not entangled at all.
\item For a fermionic Slater determinant $\Psi_N=\phi_1\wedge \phi_2\wedge\ldots \wedge \phi_N$, an elementary calculation\footnote{See e.g.\ \cite[Appendix E]{LW}} gives $S(\gam_k)=\log\binom{N}{k}$. For $k\ll N$, this gives that $S(\gam_k)\sim k\log N$ is essentially linear.
\end{itemize}

Of course, these two classes of many-body states are very special and are expected to arise in practice only as eigenstates of non-interacting Hamiltonians. Since realistic quantum many-body systems are interacting, it is an important problem in quantum information theory and quantum many-body theory to understand the behavior of $S(\gam_k)$ for more general pure states. Previous works on this topic have mainly focused on fermions, often in specific situations \cite{wolf2006violation,banuls2007entanglement,eckert2002quantum,wang2005canonical,gioev2006entanglement,kraus2009pairing,calabrese2012entanglement} (e.g., small $N$-values, free fermions, or for BCS pairing states) or from the perspective of optimization problems for $k=1,2$ \cite{Coleman,CLR,christiansen2024hilbert,yang1962concept}. For fermions, this is related to the well-known \textit{$N$-representability problem} which asks for a simple characterization of the set of possible $\gamma_2$'s and whose full solution would have immense practical ramifications for quantum chemistry \cite{Coleman,klyachko2006quantum}.

Our focus here is completely different: We are not interested in calculating $S_k$'s (which depend sensitively on the physical properties of the quantum state) but instead we identify \textit{general relations} among the entanglement entropies $S_k$ at different $k$-values. Remarkably, the relations are \textit{universal} in two key ways: (i) They only require indistinguishability and are in particular independent of the particle statistics and (ii) they are non-asymptotic, i.e., they hold for all values of $k$ and $N$, small and large. This manuscript is a streamlined, more focused version of the unpublished arXiv preprint \cite{lemm2017entropy}.

As the above references show, the study of entanglement entropy in quantum spin systems (where degrees of freedom are \textit{fixed} in space) has been widely developed in the past 20 years. This has immense practical impact because controlling the amount of entanglement of a quantum state, say, through an area law, is connected to useful approximation schemes such as matrix product states (and higher-dimensional tensor network states) \cite{verstraete2006criticality,hastings2007area,eisert2010colloquium,landau2015polynomial,anshu2022area} which lie at the heart of state-of-the-art numerical algorithms \cite{schollwock2011density}. By comparison, our understanding of quantum entanglement is substantially less developed for quantum gases of many strongly interacting and mobile particles --- especially for interacting bosons \cite{abrahamsen2022entanglement,lemm2022thermal}.  The goal of the present paper is to contribute very general, simple entropic relations that may be of use in this budding field.

\subsection{Main result}

We call an $N$-particle state permutation-invariant if it commutes with any permutation of the $N$ tensor components (i.e., if it is invariant under relabeling the particles). We remark that $\rho_N=\ket{\Psi_N}\bra{\Psi_N}$ is permutation-invariant precisely when the particles are indistinguishable. Typically, this means that the wave function $\ket{\Psi_N}$ is fermionic or bosonic, but it can also be anyonic. 

We prove the following entropic entanglement relations.
\be{thm}[Main result]
\label{thm:main1}
Let $\rho_N=\ket{\Psi_N}\bra{\Psi_N}$ be a permutation-invariant pure state on the $N$-particle Hilbert space $\mathcal H^{\otimes N}$. The map $k\mapsto S_k$ has the following properties.
\be{enumerate}[label=(\roman*)]
\item \textbf{Monotonicity.} For every $1\leq k\leq \frac{N-1}{2}$,
\beq\label{eq:main11}
S_k\leq S_{k+1}.
\eeq

\item \textbf{Concavity.} For every $2\leq k\leq N-1$,
\beq\label{eq:main12}
S_k\geq \frac{S_{k+1}+S_{k-1}}{2}.
\eeq
\e{enumerate}
\e{thm}

%
%
%
%
%
%
The monotonicity \eqref{eq:main11} is the more surprising property, since for general non-permutation-invariant quantum states, the von Neumann entropy is not at all monotonic under the partial trace. For example, consider the case $k=1$. Suppose the local Hilbert space is $d$-dimensional, $\mathcal H=\mathbb C^d$, and $N$ is even. Consider the following \textit{non}-permutation-invariant state which is constructed by taking $N/2$ tensor copies of the so-called maximally entangled state $\ket{\Xi_2}$, i.e.,
\[
\ket{\Psi_N}=\bigotimes_{k=1}^{N/2}\ket{\Xi_2},\qquad \ket{\Xi_2}=\frac{1}{\sqrt d} \sum_{j=1}^d \ket{j\otimes j}.
\]
For this $\ket{\Psi_N}$, $\gamma_2=\ket{\Xi_2}\bra{\Xi_2}$ is a pure state and thus has entropy $S_2=0$, while $\gamma_1=\frac{1}{d}\mathrm{Id}_{\mathbb C^d}$ has entropy $S_1=\log d$, which is maximal. That is, without permutation-invariance, monotonicity can fail in a strong way. 

Theorem \ref{thm:main1} establishes that monotonicity does hold for  the family of reduced $k$-body density matrices $\{\gamma_k\}_{1\leq k<N/2}$ of indistinguishable particles, which is therefore a family of quantum states with significant additional structure. (For example, symmetrizing the product of maximally entangled states from above restores the monotonicity.)

Monotonicity also implies  that
\[
\argmax\limits_{1\leq k\leq N} S_k=\l\lfloor\frac{N}{2}\r\rfloor,
\]
in accordance with the intuition that entanglement is maximal if the particles are split into (almost) equal-size groups.

Regarding concavity, notice that it can be rephrased as $S_{k+1}-S_k\leq S_{k}-S_{k-1}$. As this rewriting suggests, concavity can be viewed as a version of strong subadditivity of entropy \cite{LiebRuskai} under additional constraints coming from the many-body indistinguishability.\\

As a rule-of-thumb, these relations are not informative by themselves but rather in conjunction with explicit entropy calculations or entropic bounds that take into account further information about the many-body system under investigation. The goal of this short note is to present the very general entropic relations and we leave concrete applications to future work. As a very brief example of an application that holds for general fermionic pure states, we combine our entropic relations with a recent general bound $S(\gamma_2)\geq \log \frac{(N-1)^2}{5}$ for fermionic pure states \cite{christiansen2024hilbert}.

\be{cor}[Lower bound on entanglement entropy of fermions for all $k$]\label{cor:appl}
Suppose that $\ket{\Psi_N}$ is a fermionic $N$-particle state on $\mathcal H^{\otimes N}$ and let $2\leq k\leq N-2$. We have the lower bound on the entanglement entropy of $\gamma_k$,
\beq\label{eq:vNapplication}
S(\gamma_k)\geq \log \frac{(N-1)^2}{5}=2\log N -\log 5+o(1), \qquad \textnormal{as } N\to\infty.
\eeq
\e{cor}

The corollary gives a lower bound on the entanglement between $k$ fermions with the remaining $N-k$ ones in arbitrary quantum states. To prove Corollary \ref{cor:appl} we simply note that \cite[Eq. (1.13)]{christiansen2024hilbert} proves $S(\gamma_2)\geq \log \frac{(N-1)^2}{5}$ and then we apply the monotonicity part of Theorem \ref{thm:main1} as well as the symmetry $S_k=S_{N-k}$. 

Also for $k>2$, it is expected expect that $S(\gamma_k)\sim k\log N$ for large $N$ since Slater determinants are believed to be minimally entangled among fermionic states, cf.\ \cite{CLR}, but the bound we obtain in \eqref{eq:vNapplication} above is currently the best available one. On a related note, it is worth pointing out that the concavity part of our result says that it would be enough to prove $S(\gamma_k)\gtrsim k\log N$ for either even or odd $k$ only (e.g., knowing $S(\gam_4)\gtrsim 4\log N$ would imply $S(\gam_3)\gtrsim 3\log N$ by concavity).

We prove Theorem \ref{thm:main1} using basic tools from quantum information entropy. The idea is to consider \textit{relative} entropies of suitably chosen subsystems and to apply the data processing inequality for a certain partial trace operation. The (very short) proof is reminscient of the way strong subadditivity of entropy follows from monotonicity of the partial trace, but it also leverages indistinguishability in a key way. 

Regarding possible extensions, we note that the proof uses properties of the logarithm and does not extend to Renyi entropies in an obvious way. The data processing inequality for the standard (Umegaki) relative entropy has been refined in recent years by identifying a remainder term related to recoverability in terms of (rotated) Petz channels \cite{fawzi2015quantum,berta2015monotonicity,sutter2016strengthened,sutter2017multivariate,junge2018universal,lemm2018multivariate,gao2021recoverability}. It is straightforward to employ these refinements in the arguments below to obtain refined versions of Theorem \ref{thm:main1}; see also \cite{LW}. For simplicity of the presentation, we have not opted for this in the present manuscript.  It would be interesting to extend the ideas presented here to multipartite entanglement and/or to mixed state entanglement.

\section{Proof}

\subsection{Quantum relative entropy}
As mentioned in the introduction, they leverage the symmetry property $S(\gam_k)=S(\gam_{N-k})$ and the data processing inequality for the quantum relative entropy (specifically its monotonicity under the partial trace channel), which we recall now.
\be{defn}[Relative entropy]
Given two quantum states $\rho$ and $\sigma$, their quantum relative entropy is defined by
$$
\begin{aligned}
D(\rho\|\sigma):=
\begin{cases}
\Tr[\rho(\log\rho-\log\sigma)], \qquad &\textnormal{if } \ker\sigma\subset\ker \rho,\\
\infty,\qquad &\textnormal{otherwise.}
\end{cases}
\end{aligned}
$$
\e{defn}

The data processing inequality \cite{LiebRuskai,NC} implies that the relative entropy decreases under application of the partial trace. Namely, if $\rho_{AB},\sigma_{AB}$
 are quantum states on a Hilbert space $\curly{H}_A\otimes \curly{H}_B$, then
\beq\label{eq:mono}
 D(\rho_{AB}\|\sigma_{AB})\geq D(\Tr_B[\rho_{AB}]\|\Tr_B[\sigma_{AB}]).
\eeq

Note that the quantum relative entropy may be infinite if the supports of the states do not match suitably. To exclude this possibility in the applications below, we will use the following lemma.

\be{lm}\label{lm:kernel}
Let let $k_1,k_2\geq 1$ with $k_1+k_2\leq N$. Then
\[
\ker (\gamma_{k_1}\otimes \gamma_{k_2})\subset \ker \gamma_{k_1+k_2}.
\]
\e{lm}

\be{proof}[Proof of Lemma \ref{lm:kernel}]
By the spectral theorem, we can diagonalize $\gamma_k$ for every $k$,
\[
\gamma_k=\sum_{j\geq 1} \lambda^{(k)}_j \ket{\alpha^{(k)}_j}\bra{\alpha^{(k)}_j}.
\]
Then 
\[
\ker (\gamma_{k_1}\otimes \gamma_{k_2})=\mathrm{span}\setof{\ket{\alpha^{(k_1)}_{j_1}\otimes \alpha^{(k_2)}_{j_2}}}{\lambda^{(k_1)}_{j_1}\lambda^{(k_2)}_{j_2}=0}.
\]
Consider a fixed element of the basis of $\ker (\gamma_{k_1}\otimes \gamma_{k_2})$, i.e., a vector $\ket{\alpha^{(k_1)}_{j_1}\otimes \alpha^{(k_2)}_{j_2}}$ and assume without loss of generality that $\lambda^{(k_1)}_{j_1}=0$. Then we have
\[
\begin{aligned}
&\bra{\alpha^{(k_1)}_{j_1}\otimes \alpha^{(k_2)}_{j_2}} \gamma_{k_1+k_2}
\ket{\alpha^{(k_1)}_{j_1}\otimes \alpha^{(k_2)}_{j_2}}\\
=&\Tr[ \gamma_{k_1+k_2}
\ket{\alpha^{(k_1)}_{j_1}\otimes \alpha^{(k_2)}_{j_2}}\bra{\alpha^{(k_1)}_{j_1}\otimes \alpha^{(k_2)}_{j_2}}]\\
=&\Tr[ \sqrt{\gamma_{k_1+k_2}}
\ket{\alpha^{(k_1)}_{j_1}\otimes \alpha^{(k_2)}_{j_2}}\bra{\alpha^{(k_1)}_{j_1}\otimes \alpha^{(k_2)}_{j_2}} \sqrt{\gamma_{k_1+k_2}}]\\
\leq&\Tr[ \sqrt{\gamma_{k_1+k_2}}
(\ket{\alpha^{(k_1)}_{j_1}}\bra{\alpha^{(k_1)}_{j_1}}\otimes \mathbbm 1) \sqrt{\gamma_{k_1+k_2}}]\\
=&\Tr[(\ket{\alpha^{(k_1)}_{j_1}}\bra{\alpha^{(k_1)}_{j_1}}\otimes \mathbbm 1)\gamma_{k_1+k_2}]\\
=&\Tr[\ket{\alpha^{(k_1)}_{j_1}}\bra{\alpha^{(k_1)}_{j_1}}\gamma_{k_1}]\\
=&\lambda^{(k_1)}_{j_1}\\
=&0,
\end{aligned}
\]
where we used the definition of the partial trace in the third-to-last step.
\e{proof}
 We will use Lemma \ref{lm:kernel} several times in the arguments below without making explicit reference to it every time.

\subsection{Proof of Theorem \ref{thm:main1}}
We begin with the concavity estimate \eqref{eq:main12}, since it is slightly easier. Let $2\leq k\leq N-1$. By permutation-invariance and the data processing inequality \eqref{eq:mono}, we have
\beq\label{eq:conc}
\begin{aligned}
&D(\gam_{k+1}\|\gam_1\otimes \gam_k)-D(\gam_k\|\gam_1\otimes \gam_{k-1})\\
=&D(\gam_{k+1}\|\gam_1\otimes \gam_k)-D(\Tr_{k+1}[\gam_{k+1}]\|\Tr_{k+1}[\gam_1\otimes \gam_k])\\
\geq& 0.
\end{aligned}
\eeq
Using that $\log(X_A\otimes Y_B)=\log X_A\otimes I_B+I_A\otimes \log Y_B$ and recalling the definition of the partial trace, we can express the left-hand side in terms of $S_{k-1},S_k$ and $S_{k+1}$ as follows.
$$
\begin{aligned}
D(\gam_{k+1}\|\gam_1\otimes \gam_k)
=&\Tr[\gam_{k+1}\log \gam_{k+1}]-\Tr[\gam_{k+1} (\log \gam_1 \otimes I_{(\C^d)^{\otimes k}} +I_{\C^d}\otimes \log\gam_k)]\\
=&-S_{k+1}-\Tr[\gam_1 \log\gam_1]-\Tr[\gam_k\log\gam_k]\\
=&-S_{k+1}+S_1+S_k.
\end{aligned}
$$
Applying this identity twice to \eqref{eq:conc}, we get $-S_{k+1}+S_1+S_k-(-S_k+S_1+S_{k-1})\geq 0
$ and this is equivalent to the concavity \eqref{eq:main12}.

Next we prove the monotonicity \eqref{eq:main11}. Let $1\leq k\leq \frac{N-1}{2}$, so that $N-2k-1\geq 0$. The right quantity to apply the data processing inequality to is now
$$
\begin{aligned}
&D(\gam_{N-k}\|\gam_1\otimes \gam_{N-k-1})-D(\gam_{k+1}\|\gam_1\otimes \gam_k)\\
=&D(\gam_{N-k}\|\gam_1\otimes \gam_{N-k-1})-D(\Tr_{k+2,\ldots,N-k}[\gam_{N-k}]\|\Tr_{k+2,\ldots,N-k}[\gam_1\otimes \gam_{N-k-1}])\\
\geq& 0.
\end{aligned}
$$
Here we used the convention that $\Tr_{k+2,\ldots,N-k}[X]=X$ if $N-2k-1=0$. Using the symmetry $S_k=S_{N-k}$, we find
$$
D(\gam_{N-k}\|\gam_1\otimes \gam_{N-k-1})=-S_{N-k}+S_1+S_{N-k-1}=-S_k+S_1+S_{k+1}.
$$
Therefore, we have $-S_k+S_1+S_{k+1}-(-S_{k+1}+S_1+S_{k})\geq 0$ which is equivalent to $S_{k+1}\geq S_k$, i.e., \eqref{eq:main11}.
This concludes the proof of Theorem \ref{thm:main1}.
\qed

\section*{Acknowledgments}
It is a pleasure to thank Rupert L.\ Frank and Elliott H.\ Lieb for helpful discussions and encouragement.


\begin{thebibliography}{100}


\bibitem{bennett1992communication}
Charles~H Bennett and Stephen~J Wiesner.
\newblock Communication via one-and two-particle operators on
  einstein-podolsky-rosen states.
\newblock {\em Physical review letters}, 69(20):2881, 1992.

\bibitem{bennett1996purification}
Charles~H Bennett, Gilles Brassard, Sandu Popescu, Benjamin Schumacher, John~A
  Smolin, and William~K Wootters.
\newblock Purification of noisy entanglement and faithful teleportation via
  noisy channels.
\newblock {\em Physical review letters}, 76(5):722, 1996.

\bibitem{bennett1996mixed}
Charles~H Bennett, David~P DiVincenzo, John~A Smolin, and William~K Wootters.
\newblock Mixed-state entanglement and quantum error correction.
\newblock {\em Physical Review A}, 54(5):3824, 1996.

\bibitem{horodecki2009quantum}
Ryszard Horodecki, Pawel Horodecki, Michal Horodecki, and Karol Horodecki.
\newblock Quantum entanglement.
\newblock {\em Reviews of modern physics}, 81(2):865, 2009.

\bibitem{buscemi2011entanglement}
Francesco Buscemi and Nilanjana Datta.
\newblock Entanglement cost in practical scenarios.
\newblock {\em Physical review letters}, 106(13):130503, 2011.

\bibitem{datta2014second}
Nilanjana Datta and Felix Leditzky.
\newblock Second-order asymptotics for source coding, dense coding, and
  pure-state entanglement conversions.
\newblock {\em IEEE Transactions on Information Theory}, 61(1):582--608, 2014.

\bibitem{datta2016second}
Nilanjana Datta, Marco Tomamichel, and Mark~M Wilde.
\newblock On the second-order asymptotics for entanglement-assisted
  communication.
\newblock {\em Quantum Information Processing}, 15(6):2569--2591, 2016.

\bibitem{NC}
Michael~A Nielsen and Isaac~L Chuang.
\newblock Quantum computation and quantum information.
\newblock {\em Phys. Today}, 54(2):60, 2001.

\bibitem{calabrese2004entanglement}
Pasquale Calabrese and John Cardy.
\newblock Entanglement entropy and quantum field theory.
\newblock {\em Journal of statistical mechanics: theory and experiment},
  2004(06):P06002, 2004.

\bibitem{kitaev2006topological}
Alexei Kitaev and John Preskill.
\newblock Topological entanglement entropy.
\newblock {\em Physical review letters}, 96(11):110404, 2006.

\bibitem{hayden2006aspects}
Patrick Hayden, Debbie~W Leung, and Andreas Winter.
\newblock Aspects of generic entanglement.
\newblock {\em Communications in mathematical physics}, 265:95--117, 2006.

\bibitem{verstraete2006criticality}
Frank Verstraete, Michael~M Wolf, David Perez-Garcia, and J~Ignacio Cirac.
\newblock Criticality, the area law, and the computational power of projected
  entangled pair states.
\newblock {\em Physical review letters}, 96(22):220601, 2006.

\bibitem{banuls2007entanglement}
Mari-Carmen Banuls, Juan~Ignacio Cirac, and Michael~M Wolf.
\newblock Entanglement in fermionic systems.
\newblock {\em Physical Review A}, 76(2):022311, 2007.

\bibitem{hastings2007area}
Matthew~B Hastings.
\newblock An area law for one-dimensional quantum systems.
\newblock {\em Journal of statistical mechanics: theory and experiment},
  2007(08):P08024, 2007.

\bibitem{amico2008entanglement}
Luigi Amico, Rosario Fazio, Andreas Osterloh, and Vlatko Vedral.
\newblock Entanglement in many-body systems.
\newblock {\em Reviews of modern physics}, 80(2):517, 2008.

\bibitem{brandao2008entanglement}
Fernando~GSL Brandao and Martin~B Plenio.
\newblock Entanglement theory and the second law of thermodynamics.
\newblock {\em Nature Physics}, 4(11):873--877, 2008.

\bibitem{nishioka2009holographic}
Tatsuma Nishioka, Shinsei Ryu, and Tadashi Takayanagi.
\newblock Holographic entanglement entropy: an overview.
\newblock {\em Journal of Physics A: Mathematical and Theoretical},
  42(50):504008, 2009.

\bibitem{eisert2010colloquium}
Jens Eisert, Marcus Cramer, and Martin~B Plenio.
\newblock Colloquium: Area laws for the entanglement entropy.
\newblock {\em Reviews of modern physics}, 82(1):277, 2010.

\bibitem{christandl2012entanglement}
Matthias Christandl, Norbert Schuch, and Andreas Winter.
\newblock Entanglement of the antisymmetric state.
\newblock {\em Communications in Mathematical Physics}, 311(2):397--422, 2012.

\bibitem{brandao2015exponential}
Fernando~GSL Brandao and Michal Horodecki.
\newblock Exponential decay of correlations implies area law.
\newblock {\em Communications in mathematical physics}, 333:761--798, 2015.

\bibitem{landau2015polynomial}
Zeph Landau, Umesh Vazirani, and Thomas Vidick.
\newblock A polynomial time algorithm for the ground state of one-dimensional
  gapped local {H}amiltonians.
\newblock {\em Nature Physics}, 11(7):566--569, 2015.

\bibitem{gong2017entanglement}
Zhe-Xuan Gong, Michael Foss-Feig, Fernando~GSL Brand{\~a}o, and Alexey~V
  Gorshkov.
\newblock Entanglement area laws for long-range interacting systems.
\newblock {\em Physical review letters}, 119(5):050501, 2017.

\bibitem{cirac2021matrix}
J~Ignacio Cirac, David Perez-Garcia, Norbert Schuch, and Frank Verstraete.
\newblock Matrix product states and projected entangled pair states: Concepts,
  symmetries, theorems.
\newblock {\em Reviews of Modern Physics}, 93(4):045003, 2021.

\bibitem{anshu2022area}
Anurag Anshu, Itai Arad, and David Gosset.
\newblock An area law for 2d frustration-free spin systems.
\newblock In {\em Proceedings of the 54th Annual ACM SIGACT Symposium on Theory
  of Computing}, pages 12--18, 2022.

\bibitem{jauslin2022random}
Ian Jauslin and Marius Lemm.
\newblock Random translation-invariant hamiltonians and their spectral gaps.
\newblock {\em Quantum}, 6:790, 2022.

\bibitem{islam2015measuring}
Rajibul Islam, Ruichao Ma, Philipp~M Preiss, M~Eric~Tai, Alexander Lukin,
  Matthew Rispoli, and Markus Greiner.
\newblock Measuring entanglement entropy in a quantum many-body system.
\newblock {\em Nature}, 528(7580):77--83, 2015.

\bibitem{LW}
Marius Lemm and Mark~M Wilde.
\newblock Information-theoretic limitations on approximate quantum cloning and
  broadcasting.
\newblock {\em Physical Review A}, 96(1):012304, 2017.

\bibitem{wolf2006violation}
Michael~M Wolf.
\newblock Violation of the entropic area law for fermions.
\newblock {\em Physical review letters}, 96(1):010404, 2006.

\bibitem{eckert2002quantum}
K~Eckert, John Schliemann, D~Bru{\ss}, and M~Lewenstein.
\newblock Quantum correlations in systems of indistinguishable particles.
\newblock {\em Annals of physics}, 299(1):88--127, 2002.

\bibitem{wang2005canonical}
Xiao-Guang Wang and Barry~C Sanders.
\newblock Canonical entanglement for two indistinguishable particles.
\newblock {\em Journal of Physics A: Mathematical and General}, 38(5):L67,
  2005.

\bibitem{gioev2006entanglement}
Dimitri Gioev and Israel Klich.
\newblock Entanglement entropy of fermions in any dimension and the widom
  conjecture.
\newblock {\em Physical review letters}, 96(10):100503, 2006.

\bibitem{kraus2009pairing}
Christina~V Kraus, Michael~M Wolf, J~Ignacio Cirac, and G{\'e}za Giedke.
\newblock Pairing in fermionic systems: A quantum-information perspective.
\newblock {\em Physical Review A}, 79(1):012306, 2009.

\bibitem{calabrese2012entanglement}
Pasquale Calabrese, Mihail Mintchev, and Ettore Vicari.
\newblock Entanglement entropies in free-fermion gases for arbitrary dimension.
\newblock {\em Europhysics Letters}, 97(2):20009, 2012.

\bibitem{Coleman}
John Coleman.
\newblock Structure of fermion density matrices.
\newblock {\em Reviews of modern Physics}, 35(3):668, 1963.

\bibitem{CLR}
Eric~A Carlen, Elliott~H Lieb, and Robin Reuvers.
\newblock Entropy and entanglement bounds for reduced density matrices of
  fermionic states.
\newblock {\em Communications in mathematical physics}, 344(3):655--671, 2016.

\bibitem{christiansen2024hilbert}
Martin~Ravn Christiansen.
\newblock Hilbert--schmidt estimates for fermionic 2-body operators.
\newblock {\em Communications in Mathematical Physics}, 405(1):1--9, 2024.

\bibitem{yang1962concept}
Chen~Ning Yang.
\newblock Concept of off-diagonal long-range order and the quantum phases of
  liquid {H}e and of superconductors.
\newblock {\em Reviews of Modern Physics}, 34(4):694, 1962.

\bibitem{klyachko2006quantum}
Alexander~A Klyachko.
\newblock Quantum marginal problem and {N}-representability.
\newblock In {\em Journal of Physics: Conference Series}, volume~36, page~72.
  IOP Publishing, 2006.

\bibitem{lemm2017entropy}
Marius Lemm.
\newblock On the entropy of fermionic reduced density matrices.
\newblock {\em arXiv preprint arXiv:1702.02360 [v1]}, 2017.

\bibitem{schollwock2011density}
Ulrich Schollw{\"o}ck.
\newblock The density-matrix renormalization group in the age of matrix product
  states.
\newblock {\em Annals of physics}, 326(1):96--192, 2011.

\bibitem{abrahamsen2022entanglement}
Nilin Abrahamsen, Yu~Tong, Ning Bao, Yuan Su, and Nathan Wiebe.
\newblock Entanglement area law for 1d gauge theories and bosonic systems.
\newblock {\em arXiv preprint arXiv:2203.16012}, 2022.

\bibitem{lemm2022thermal}
Marius Lemm and Oliver Siebert.
\newblock Thermal area law for lattice bosons.
\newblock {\em arXiv preprint arXiv:2207.07760}, 2022.

\bibitem{LiebRuskai}
Elliott~H Lieb and Mary~Beth Ruskai.
\newblock A fundamental property of quantum-mechanical entropy.
\newblock {\em Physical Review Letters}, 30(10):434, 1973.

\bibitem{fawzi2015quantum}
Omar Fawzi and Renato Renner.
\newblock Quantum conditional mutual information and approximate {M}arkov
  chains.
\newblock {\em Communications in Mathematical Physics}, 340(2):575--611, 2015.

\bibitem{berta2015monotonicity}
Mario Berta, Marius Lemm, and Mark~M Wilde.
\newblock Monotonicity of quantum relative entropy and recoverability.
\newblock {\em Quantum Information \& Computation}, 15(15-16):1333--1354, 2015.

\bibitem{sutter2016strengthened}
David Sutter, Marco Tomamichel, and Aram~W Harrow.
\newblock Strengthened monotonicity of relative entropy via pinched {P}etz
  recovery map.
\newblock {\em IEEE Transactions on Information Theory}, 62(5):2907--2913,
  2016.

\bibitem{sutter2017multivariate}
David Sutter, Mario Berta, and Marco Tomamichel.
\newblock Multivariate trace inequalities.
\newblock {\em Communications in Mathematical Physics}, 352:37--58, 2017.

\bibitem{junge2018universal}
Marius Junge, Renato Renner, David Sutter, Mark~M Wilde, and Andreas Winter.
\newblock Universal recovery maps and approximate sufficiency of quantum
  relative entropy.
\newblock In {\em Annales Henri Poincar{\'e}}, volume~19, pages 2955--2978.
  Springer, 2018.

\bibitem{lemm2018multivariate}
Marius Lemm.
\newblock {On multivariate trace inequalities of {S}utter, {B}erta, and
  {T}omamichel}.
\newblock {\em Journal of Mathematical Physics}, 59(1), 2018.

\bibitem{gao2021recoverability}
Li~Gao and Mark~M Wilde.
\newblock Recoverability for optimized quantum f-divergences.
\newblock {\em Journal of Physics A: Mathematical and Theoretical},
  54(38):385302, 2021.

\end{thebibliography}
%
\begin{footnotesize}

\end{footnotesize}

\end{document}